\documentstyle[11pt,paspconf,rotate,epsf,times]{article}

\def\xvv{\hbox{$15\!-\!V$}}
\def\ngc#1{\hbox{\rm NGC\thinspace#1}}
\def\et{et~al.}
\def\teff{\mathmode{T_{_{\rm eff}}}}
\def\mathmode#1{\ifmmode {#1} 
                  \else {$#1\mkern-5mu$} \fi}

\def\fh{\mathmode{f_{_{\rm H}}}}

\def\msun{\mathmode{M_\odot}}
\def\mlam{\mathmode{m_\lambda}}
\def\feh{{\rm [Fe/H]}}

\begin{document}

\vbox{\noindent\small\it To appear in `From Stars to Galaxies: The Impact
  of Stellar Physics on Galaxy Evolution,' eds C. Leitherer, U. Fritze-von
  Alvensleben, \& J. P. Huchra, ASP Conf Ser}

\title{ Constraints on the Stellar Populations of Elliptical Galaxies
from Ultraviolet Spectra}
\author{Ben Dorman}
\affil{Laboratory for Astronomy \& Solar Physics,\\
Code 681, NASA/GSFC,  Greenbelt, MD USA 20771}
\author{Robert W. O'Connell}
\affil{Dept of Astronomy, University of Virginia,\\
P. O. Box 3818, Charlottesville, VA USA 22903-0818}

\begin{abstract}
We present preliminary results from spectral synthesis models of old
stellar populations for the spectral range 912-4000 \AA\ with $\sim 10\, {\rm
\AA}$ resolution, which can be used to investigate the UVX phenomenon and
to assess ages and abundances.
Model spectra incorporating
extreme horizontal branch (EHB)  and Post-Asymptotic Giant Branch
(P-AGB) populations give good matches to the far-UV spectra
of galaxies.  These models indicate an EHB
fraction which is $< 10\%$ of the total HB population in all but the
most extreme examples of the UVX phenomenon, where the EHB fraction is
still $\leq 20\%.$  Once the hot component that
gives rise to the UVX phenomenon is accounted for, the mid-UV wavelength
range ($2200 < \lambda < 3300~{\rm \AA}$) provides information about
the age and metallicity of the underlying stellar population. 
The flux in this spectral range arises mainly from 
stars close to the main sequence turnoff.  We compare models with
the spectrum of M31 and discuss
UV features which should be useful as population diagnostics.

\end{abstract}


\section{Introduction}

We present the results of our continuing investigation of ultraviolet
radiation from old stellar populations. In this paper, we discuss the
far-UV upturn phenomenon and point out the advantages of the long
wavelength
end of the vacuum UV window (the so-called ``mid-UV'', $2000 \la
\lambda \la 3300\: {\rm \AA}$) for determining the age and abundance
of the bulk stellar population.  

The sources most widely held now to be responsible for the far-UV
upturn (or UVX) phenomenon are extreme horizontal branch (EHB) stars
(e.g. Burstein \et\ 1988; Greggio \& Renzini 1990; Dorman, Rood, \&
O'Connell 1993, hereafter Paper I; Dorman, O'Connell, \& Rood 1995,
hereafter Paper II, and references therein). Unfortunately, the
production of these stars occurs through mass loss processes in cool
giants, which remain  ill-understood and unpredictable with our
current understanding of stellar evolution.  Changes of only a few
0.01\msun\ in envelope mass can strongly affect the far-UV output of hot
HB stars.  In contrast, the mid-UV flux is dominated by the turnoff
population, which responds strongly to the age and abundance of the
underlying stellar population.  This region can therefore potentially
provide us with population indicators that suffer less from the
well-known degeneracies found at optical wavelengths (e.g.  Worthey
1994 and this meeting).

\begin{figure}[t]
\epsfxsize=5.25truein
{\epsffile{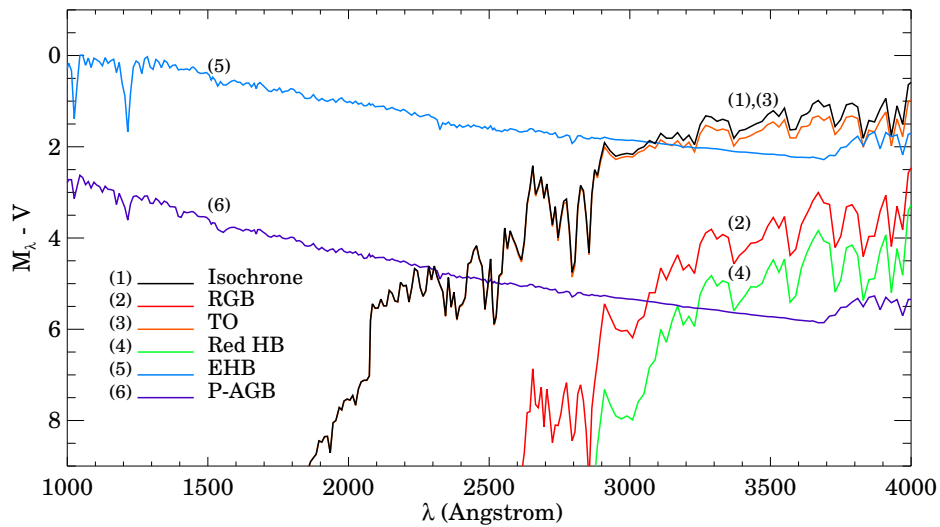}}\hfil
\caption{
\protect\label{fig:compts} \protect\small Components of an old stellar population in
the ultraviolet. Illustrated is a model with [Fe/H] $= 0$ at 10 Gyr.
The pre-He flash component is represented by three separate
lines: (1) the total light; (2) the giant branch [$\teff \le 5000\: {\rm K}, L > L_{_{\rm
TO}}$]; and (3) the turnoff [$\teff > 5000\: {\rm K}$].  The maximum possible
contributions due to red HB stars (4), EHB stars (5), and P-AGB stars
(6) are also shown (see
section~\protect\ref{sec:fuv}).}
\end{figure}

We illustrate the UV spectra of components of old metal-rich stellar
populations in Fig.\ref{fig:compts}, plotted in magnitudes normalized
at $V$.  These are based on the same evolutionary models discussed in
Paper II, but we have now added spectral energy distributions using
 full resolution model stellar fluxes
taken from the Kurucz (1993, CD-ROM) grid. The maximum possible P-AGB and EHB 
contributions are shown to the same scale; each of these would be realized
if {\it all} post-RGB stars went through that particular channel.
Note that the most obvious difference between (5)
and (6) is in the strength of the Lyman series, allowing a possible
diagnostic for the relative contributions of these two hot sources.
  
The most striking feature of this diagram is that the turnoff flux
dominates the contribution from the earlier stages of evolution at all
wavelengths shortward of 4000~\AA; the continuum produced by the cool
giants is about 2 mag below that of the turnoff (compare Buzzoni 1989 and
Worthey 1994). Thus the mid-UV spectrum
($\lambda > 2000~{\rm \AA}$) derives almost entirely from the turnoff
and from the hot component that gives rise to the UVX phenomenon
(Burstein \et\ 1988). If the spectrum emanates solely from old stars,
then the mid-UV radiation can be corrected for the UVX component to
yield important information about the bulk stellar population.

\section{\label{sec:fuv} The Far-UV (\mbox{$\bf 912 < \lambda < 2000$} \AA)}

The galaxies with the strongest manifestations of the UV upturn
phenomenon (\ngc{1399} and \ngc{4649}  $\rm = M60),$ with $\mlam(1500 {\rm
\AA}) - V = \xvv \la 2.5$ have rightly received a great deal of
attention. However in most of the ellipticals so far observed, $3 < \xvv <
4.$ The evidence gathered in the last few years by HST, UIT and HUT has
all but ruled out young massive stellar components as the explanation
for the UV spectra of these galaxies (Paper II).  In the `strong UVX'
systems the radiation also cannot be explained by the hot sources --- post
Asymptotic Giant Branch (P-AGB) stars, including planetary nebula
nuclei--- that must be present in old, passively evolving
stellar populations (Greggio \& Renzini 1990).
In the `weaker UVX' cases $(\xvv > 3.5),$ however, an explanation in terms
of the `classical' P-AGB stars is plausible given the remaining uncertainties
in the stellar models.

The stellar populations of the Galactic field are  however 
found to contain other types of UV-bright stars.
The Palomar-Green survey of UV excess sources (Green, Schmidt, \& Liebert
1986) is dominated by the Galactic subdwarf B and O stars (see Saffer \&
Liebert 1995 for a recent study), which are 
apparently produced by extreme mass loss in single stars.
These are the observed counterparts of the EHB stars and of their
post-HB descendants, the AGB-manqu\'e objects (Paper II).
While they appear to be a natural extension of the horizontal
branch sequence seen in metal-poor clusters, this may not be
the full explanation.
Both in the Galactic field and in the few globular clusters
where they have been found ($\omega$ Cen:  Whitney \et\ 1994;
\ngc{6752}: Buonanno \et\ 1986), they appear to form a distinct group
separated by a temperature gap from the blue HB stars.  This may
indicate the action of a different, more vigorous mass loss process
at work in some stars. 

The simplest explanation for the variation in the UV upturn strength is
a variation in the number of EHB stars present.  The correlation of the
UVX with metallicity (Burstein \et\ 1988 \& references therein) then
arises if more stars suffer high mass loss at high $Z.$ In order to
gauge whether this explanation is plausible, it is important to make
quantitative estimates of how many such stars need to be present as a
fraction of the entire post-red giant population.  The answer to this
question depends on two main factors: first, the UV energy radiated in
the lifetime of the putative sources, and second, how many of them are
created in unit time (the `birth rate' of HB stars).

The far-UV colors of globular clusters in which nearly all of the HB
stars contribute to the ultraviolet flux are $\xvv \sim 1.6$.  In the
bluest galaxy, $\xvv \sim 2.$  Thus if the originating stars are
similar, then number of sources radiating in the UVX galaxies is a
significant fraction of the total number of HB stars present. This
argument depends on the fact that both the lifetime radiated UV flux from
core He-burning stars (Paper I) and their `birth rate' (Paper II) are
 relatively insensitive to composition.  The inference is  that the
sources responsible for the UV flux  in the strongest cases are not
drawn from a trace (i.e. $\ll 5\%$) population,   such as the stars of
the extreme metal-rich tail of the $Z$ distribution.  
The UV flux from HB stars is bounded by the
Fuel Consumption Theorem (Tinsley 1980; Renzini \& Buzzoni 1986).  The
energy radiated in the far-UV depends on the helium burnt in the core,
which varies little with composition. To be sure, during the HB phase
hydrogen shell burning can also be significant, but if the star is hot
enough to  be a strong UV source the H-shell energy is comparable or
(usually) much less than that produced by the helium core source. This
bounds the total UV energy output from objects of similar helium core
mass (see Paper I).

In Paper II we used a simple model of post-RGB phases to estimate the
fraction \fh\ of the total HB population which is on the EHB in
various observed objects.  For objects with the strongest upturns we
obtained \hbox{${\fh \sim 0.2}.$} In other words a significant, but
not dominant, portion of the population is responsible for the UV
upturn.  With extreme composition assumptions, e.g.  $\feh = 0.7,
Y=0.46$, we found $\fh = 0.10,$ the difference arising from the more
rapid evolution (and thus higher evolutionary flux) of helium rich
models rather than from the higher metallicity.  (Note that although
the helium abundance $Y$ corresponding to high metallicities is
unknown, $Y$ is unlikely to exceed its solar value by a large
percentage, because the iron is produced by stars that produce little
helium.) 

\section{\label{sec:muv} The Mid- and Near-UV ($\bf \lambda > 2000$ \AA)}

\begin{figure}[t]
\hfil{\epsffile{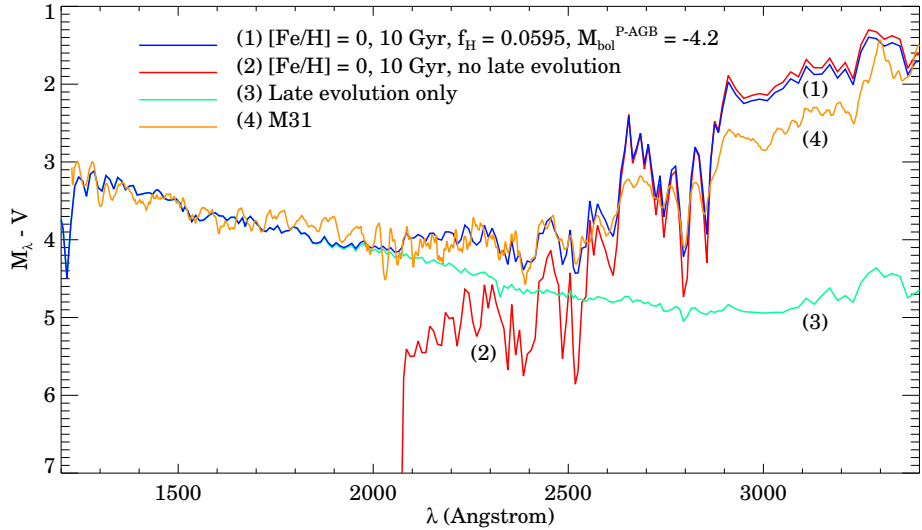}}\hfil
\caption{
\protect\label{fig:m31} \protect\small A fit to the UV/Optical spectrum
of M31 using the two component model (1) of the HB. A solar metallicity
isochrone spectrum (2) is assumed for the earlier, pre-He flash stages
(``late'' refers to the HB phase and beyond). The hot
component (3) represents a uniform distribution of EHB stars that has
been scaled to match the integrated far-UV flux from the M31 bulge,
with $\fh \sim 0.06;$ the rest of the HB clump stars are assumed to
become AGB stars and P-AGB stars. Note the match between the data and
the model at the Fe {\sc ii} feature around 2400 \AA\ and the Fe {\sc
ii} and Mg {\sc i} lines around 2500 \AA.  Also note the prominent Mg
{\sc i} and Mg {\sc ii} features at 2852 and 2800 \AA\ respectively (see
section~\protect\ref{sec:muv}).
}

\end{figure}

Figure~\ref{fig:m31} illustrates the potential utility of the longer
UV wavelengths as population diagnostics. The fraction of EHB stars in
the composite model [curve (1)] has been scaled so that its $\xvv$
color matches that of the bulge of M31 as measured by IUE.  We have
chosen M31 [curve (4)] for comparison because its spectrum has high
S/N and its far-UV flux ($\xvv \sim 3.5$) is typical of elliptical
galaxy nuclei.  The figure shows that the contributions of the turnoff
stars [curve (2)] and the hot UVX component are roughly equal for M31
at 2400--2500 \AA.  The strong lines in the turnoff component are
diluted by the UVX stars in the composite. The caption lists some
of the mid-UV spectral features that are strong in G dwarf spectra
that also appear in the galaxy spectra.  Note in particular that the
well-known Mg features around 2800 and 2850 \AA, being of different
ionization stages, may become a useful indicator of the turnoff
temperature and may provide constraints on the age and metallicity\begin{footnote}
{The Mg {\sc ii} feature may be affected by chromospheric
emission, but this is thought to be primarily a feature of younger
late-type stars: see Smith \et\ 1992}
\end{footnote}.

The turnoff spectrum has a break at 2100 \AA; the location of this
break is a strong function of metallicity (Paper III). Hence the fact
that the IUE spectrum exceeds slightly the prediction from the hot
component shortward of this point may indicate the presence of a more
metal-poor component.  Of some considerable interest is the obvious
discrepancy between the M31 spectrum and the model in the range $3000 <
\lambda < 3300~{\rm \AA}$.  We find a similar flux deficiency in this
spectral range in other galaxies (NGC 3379, NGC 4649).  Such a
deficiency is also present in the spectrum syntheses of Magris \&
Bruzual (1993), which were constructed with empirical fluxes.   This is
just at the point where IUE and ground-based datasets overlap, and the
problem may be in the calibration.  If these possibilities can be
excluded, the next most likely explanation is a localized metallic
overabundance in M31.

\section{\label{sec:future} Future Work}

The UV offers promising advantages for the study of old populations.
UV broad-band indicators, which can be used for much fainter objects,
should be developed into useful diagnostics.  At moderate redshifts ($z
\sim 1$) the mid-UV region is redshifted into the range of ground-based
optical spectroscopy.  The age dependence of the UV upturn phenomenon
is difficult to calibrate since it depends on the luminosity-age
relation of the P-AGB phase and critically on the details of giant
branch mass loss.  But one might expect that it should decrease in
intensity with lookback time since in
younger populations the P-AGB stars should be shorter-lived
and the amount of mass loss required to produce EHB stars is greater. Without the hot component the detection and
interpretation of the turnoff population at significant lookback times
becomes easier. If mid-UV spectra can be used to help separate the age
and metallicity of stellar populations, then we may be able to apply
such techniques to populations at significant redshift.  This may
ultimately do much to resolve the apparent contradiction between the
distance scale estimates of the age of the universe and those from
stellar evolution.

However, the correct interpretation of such observations will be
impossible without an adequate exploration of nearby galaxies from
space in the UV.  The observational database (e.g. Fanelli \et\ 1992;
Wu \et\ 1991) of UV spectra of the stars that dominate the turnoffs of
old stellar populations, namely late F and early G dwarfs, needs to be
improved for a well-calibrated grid of metal abundances, especially at
wavelengths below 2600 \AA.  Fig~\ref{fig:m31} also illustrates some
of the deficiencies in the Kurucz synthetic spectral atlas used to
derive the synthetic models:  for example, the sharp peak at 2600~\AA\
which is not present in the galaxy data or in IUE spectra of cool
stars. In addition a set of nearby galaxy spectra out to Lyman
$\alpha$ with higher resolution and $S/N$ than present in the IUE
atlases would be a valuable reference for conducting observational
cosmology.
 
\acknowledgments

BD is grateful for an AAS International Travel Grant which enabled his
attendance at this meeting. This work was partially supported by NASA
RTOP 188-41-51-03 and NASA Long Term Space Astrophysics Program
NAGW-2596.

\end{document}